\documentclass[twocolumn]{revtex4}
\usepackage{graphicx,amsmath,amssymb,mathrsfs}
\usepackage{epsfig}

\begin{document}

\title{Lagrangian quantum turbulence model based on alternating superfluid/normal
fluid stochastic dynamics}

\author{Shihan Miah and Christian Beck }

\affiliation{Queen Mary University of London, School of Mathematical Sciences, Mile End Road, London E1 4NS, UK}

\begin{abstract}
Inspired by recent measurements of the velocity and acceleration statistics of Lagrangian tracer
particles embedded in a turbulent quantum liquid we propose a new superstatistical model for
the dynamics of tracer particles in quantum turbulence. Our model consists of  random sequences
S/N/S/... where the particle spends some time in the superfluid (S) and some time in the normal
liquid (N). This model leads to a superposition of power law distributions generated
in the superfluid and Gaussian distributions in the normal liquid, in excellent agreement with experimental
measurements. We include memory effects into our analysis and present analytic predictions
 for probability densities and correlation functions.

\end{abstract}

\maketitle

The statistics of tracer particles embedded in a turbulent quantum liquid
is an interesting area of current turbulence research
\cite{ref11}--\cite{baggaley}. Measurements of Paoletti et al. \cite{ref11}
have shown that the measured velocity distributions asymptotically exhibit a power law with
exponent approximately given by $-3$. This is in contrast to classical fully developed turbulence,
where velocities are approximately Gaussian (but velocity {\em differences}
and accelerations are strongly non-Gaussian \cite{bodenschatz}--\cite{ref4}).
Recent measurements of La Mantia and Skrbek \cite{mantia14} have illustrated that depending on the spatial
scale probed there can be a mixture of quantum and classical features in
the distributions. The measured velocity distributions in \cite{mantia14}
again exhibit asymptotic power laws with exponent $-3$, but superimposed to that appears a Gaussian core
for low velocities. In \cite{mantia}, for the first time, also histograms of measured accelerations
for particles embedded in turbulent quantum flow were presented, which appear to be of similar shape as
those of accelerations in classical Lagrangian turbulence \cite{bodenschatz}.
Numerical simulations of the velocity statistics based
on the Gross-Pitaevskii equation have been performed in \cite{ref17}, also providing evidence for power-law tails.
A simple model for the dynamics of test particles based on $\chi^2$-superstatistics (a method of nonequilibrium
statistical mechanics \cite{beck_cohen,bcs,thurner,garcia,hasegawa})
was proposed in \cite{beck_shihan}. This model predicts power law tails
with exponent $-3$, and in addition yields further information on the shape of the probability density
near the maximum. Effectively it leads to so-called $q$-Gaussian distributions \cite{tsallis,curado,ref1} with entropic exponent $q=5/3$.

In this paper we introduce a rather general but powerful stochastic model for the dynamics
of a test particle in a turbulent quantum liquid, which, contrary to previous work,  takes into account
that there is a mixture of superfluid and normal liquid. Consequently,
this new model is based on a mixture of two different statistics: One
is related to the particle movement in the superfluid (S), leading to power law distributions, and one is based
on the movement in the normal liquid (N), leading to a superimposed Gaussian. For a given test particle
both phases alternate in a random sequence S/N/S/... If we attribute
to each symbol a fixed time scale, then repetitions of the same symbol occur as well,
e.g. SSNSNNNSS....From a statistical mechanics point of view
our model has the form of a
superstatistical stochastic differential equation \cite{ref1,ref2,ref4}
combined with a symbolic dynamics \cite{beck-schloegl,tel}.
From a condensed matter physics point of view the model has some analogy with regarding the
quantum liquid as
a spatial random collection of infinitely many S/N/S Josephson junctions \cite{dubos,prl2013}.
The symbol sequences generated by the test particle that moves through the quantum liquid
have the ability to encode complex memory effects in the dynamics
generated by the quantum turbulent flow.

We present results for the
probability distributions generated by this model and compare with recent experimental data
presented in \cite{mantia14,mantia}. Excellent agreement is found.
The model is simple enough to allow for analytical calculations of temporal correlation functions, and we
present some results for the case of a generalized dynamics that also contains a memory kernel.
So far Lagrangian correlation functions have not been measured experimentally in quantum turbulent
flow, but future measurements in this direction would be very useful. If more experimental
measurements on correlation functions were available this could help
to further narrow down the most suitable class of
stochastic models, and to better understand the symbolic dynamics generated by quantum turbulent flow.

Consider a tracer particle embedded in a quantum liquid that consists of a mixture of superfluid (S) and normal
liquid (N).
Consider a sequence (e.g. SNSNNSSSSN.....) where the local dynamics of the particle
is different
depending whether it is surrounded by liquid in phase S or N. In phase S we consider a superstatistical local dynamics
\cite{beck_shihan,ref1}
where the velocity $\boldsymbol{v}$ of the tracer particle satisfies
\begin{equation}
 \dot {\boldsymbol{v}}=-\gamma(t)\boldsymbol{v}+\omega \big[\boldsymbol e(t) \times \boldsymbol v\big] +\sigma(t) \boldsymbol{L}(t)
\label{eq1}
\end{equation}
Here $\boldsymbol{L}(t)$ is vector-valued Gaussian white noise.
We assume that the effective damping constant $\gamma$ and the noise strength $\sigma$ are functions of $t$, and so
 is $\omega$ and the direction of the unit vector $\boldsymbol{e}$,
 which is uniformly distributed. The second term on the right hand side of eq.(\ref{eq2}) represent
the rotational movement of the particle around the nearest vortex filament
in direction $\boldsymbol{e}$. The unit vector $\boldsymbol{ e}$ and the noise
strength $\sigma$ evolve stochastically on a large time scale $T_{\boldsymbol e}$ and $T_\sigma$ respectively:
 In the superstatistics approach, one consider the
{\em parameters} of a local stochastic differential equation to be
random variables \cite{ref1}. That is, the parameters in eq.~(\ref{eq1}) can take on very different values during time evolution.
A very small $\gamma$ corresponds to nearly undamped motion for a limited amount of time.
A very small $\omega$ corresponds to almost no rotation, i.e. straight movement for a limited
 amount of time. All these cases are included as possible local dynamics  and averaged over in the superstatistical approach.
Due to the integration over probability densities in $\gamma, \omega, \sigma$, eq.~(\ref{eq1}) yields non-Gaussian behavior
represented by some non-Gaussian density function $p_S(\boldsymbol{v})$
(see later sections for concrete calculations).

 Contrary to that, if the particle is surrounded by normal liquid (phase N), then we assume the dynamics is
 described by an ordinary
 Langevin equation with constant parameters $\gamma_0$ and $\sigma_0$, leading to standard type of Gaussian behavior, as expected
 for the velocity of a test particle in classical liquids:
  \begin{equation}
 \dot {\boldsymbol{v}}=-\gamma_0\boldsymbol{v}+\omega \big[\boldsymbol e(t) \times \boldsymbol v\big] +\sigma_0 \boldsymbol{L}(t)
\label{eq2}
\end{equation}
This simple linear equation just yields Gaussian behaviour with inverse variance
parameter $\beta_0=2\gamma_0/\sigma_0^2$ during phase N.
If the variance is rescaled to 1, meaning we consider the velocity in units of
its variance, then the stationary probability density in this case is simply
\begin{equation}
p_N(\boldsymbol{v})= \frac{1}{\sqrt{2\pi}}e^{-\frac{1}{2}v^2}. \label{gauss}
\end{equation}

In our model we assume that {\em both} phases are relevant
and for a given tracer particle occur in random (but not necessarily
uncorrelated) sequences.
Each tracer particle in the quantum liquid produces a temporal
sequence of symbols S and N, which, depending on parameters of
the quantum
turbulent flow environment, the size of the test particle, and the time scales involved, will
have different stochastic properties.

The simplest case is just statistically independent random sequences of symbols S an N.
The probability density $p(\boldsymbol{v})$ of velocity of the tracer particle is then given in the mixed form
\begin{equation}
p(\boldsymbol{v})=w_S p_S(\boldsymbol{v}) +w_N p_N (\boldsymbol{v})
\end{equation}
where $w_S$ and $w_N$ describe the probability of the symbol $S$, respectively $N$,
to occur $(w_s+w_N=1)$. In other words, the relative duration of the two different phases
is relevant. The probabilities $w_S$ and $w_N$
will depend on external parameters of the quantum liquid, as well as
on the spatial and temporal scale on which the measurements are done.
If the spatial scale probed by the test particle is larger than the average distance of vortex filaments, then we
expect $w_N\approx 1$ to dominate, leading basically just to Gaussian behaviour.
Nontrivial quantum turbulence properties are probed by very small test particles at very
low temperature, leading to $w_S \approx 1$.

The assumption of independent random sequences of symbols will usually be too simple for a quantum liquid.
For example, if the particle is trapped near a vortex core, then long-lasting sequences SSSSS... will occur,
similar as the laminar phase for intermittent chaotic maps near a tangent bifurcation \cite{beck-schloegl}.
We propose to condition the probabilities $w_S$ and $w_N$ on the actual velocity
of the test particle observed.
This means we may consider conditioned probabilities $w_S|_v$
and $w_N|_v$, conditioned on a given observation of the velocity $v$, in the sense that given a
large observed velocity $v$, $w_S|_v$ is close to 1, and in case a small
velocity $v$ is observed, $w_N|_v$ conditioned on that velocity is close to 1.
This is plausible, because given a large velocity $v$ it is very likely that this
happened during a phase where the particle was being embedded by the superfluid
and close to a vortex filament with rapid rotation.
In this latter model, which has strong correlations between the
observed value of $v$ and the quantities $w_S$ and $w_N$ conditioned on the
observed velocity,
the probability density is given in good approximation by
\begin{equation}
p(\boldsymbol{v})=\left\{ \begin{array}{ll}
 p_N(\boldsymbol{v}) & |v|\leq v_c \\
 p_S(\boldsymbol{v}) & |v|> v_c,
\end{array} \right. \label{max}
\end{equation}
because large velocities are almost sure to occur in phase S, and hence
above a given threshold are distributed according to the tails given by $p_S(v)$.
   $v_c$ is a critical velocity, whose value depends on the relative
occurrence probabilities $w_N$ and $w_S$: The larger $w_S$, the smaller $v_c$.
For $w_S \to 1$, $v_c \to 0$. For $w_N \to 1$, $v_c\to \infty$.

Let us now work out more details on $p_S(\boldsymbol{v})$ in phase S. 
As in the experiments, we restrict ourselves to the statistics $p_S(v_i)$
of a single component $v_i$. For ease of notation, we suppress the index $i$
in the following. We essentially follow
the approach of \cite{beck_shihan}.
Far away from
a vortex filament, the typical velocity will behave as in classical turbulence,
whereas close to a vortex filament the movement will
be very rapid and almost friction free, as superfluids imply due to
quantum mechanical constraints $v \sim 1/r$, where $r$ is the
distance of the particle to the nearest vortex core.
Following the same argument as in in ref. \cite{beck_shihan},
we consider an effective  friction dependent on the distance
to the nearest vortex filament. One obtains
for  a $\chi^2$-distributed effective friction $\gamma (t)$  of $n$ degrees of freedom (and under the assumption of a uniform
distribution of rotation axis vectors $\vec{e}$)
\begin{equation}
 p_S(v)
=\frac{\Gamma(\frac{n}{2}+\frac{1}{2})}{\Gamma(\frac{n}{2})} \left(\frac{\beta_0}{\pi n}\right)^\frac{1}{2} \frac{1}{\left(1+\frac{\beta_0}{n} v^2\right)^{\frac{n}{2}+\frac{1}{2}}} .
\label{eq16}
\end{equation}
The typical velocity of the tracer particle in the S phase depends on the perpendicular distance between the particle and the
nearest evolving (and sometimes merging) vortex filament. Therefore, the relevant degrees of freedom are $n=2$
for 3-dimensional quantum turbulence, since the distance
to a 1-dimensional vortex line has two components. In experiments there is often a drift velocity in the system
that gives a non-zero mean velocity $c$ to the test particle. In this case one has to replace $v$ by $v-c$ in  Eq.(\ref{eq16})
and (\ref{gauss}),
and for $n=2$ one ends up with
\begin{equation}
 p_S(v)=\frac{\sqrt{\beta_0}}{\left(2+\beta_0 (v-c)^2\right)^\frac{3}{2}}
\label{eq188}
\end{equation}
Clearly, for large $v$ this implies power-law tails
\begin{equation}
p_S(v)\propto {v}^{-3}.
 \end{equation}
For the above distribution $p_S$, the variance does not exist, so in practice one introduces a cutoff $v_{max}$,
so that the variance is well-defined \cite{beck_shihan}. This cutoff is also physically motivated,
one typically observes in experiments values of the velocity up to about 15 standard deviations \cite{mantia14}.

 Let us now compare our model predictions with
 recent experimental measurements performed by La Mantia \textit{et al.} \cite{mantia14,mantia}.
The experimental data obtained in \cite{mantia} are to a certain extent different from
previous seminal measurements published by Paoletti et al. in \cite{ref11}
due to the fact that the investigated flows are different. La Mantia \textit{et al.} conducted
their experiment on
 steady-state thermal counterflow, while in \cite{ref11} it was mainly decaying thermal counterflow.
Moreover, the used
techniques in both experiments were different, for example, different tracer particles were employed and the probed heat
flux range was also not the same. Therefore, one expects similar results
for the measured $p(v)$ rather than exactly the same results.
Our model can describe the experimental data obtained by both groups, assuming that for the experiment \cite{ref11} $w_N \approx 0$
(see \cite{beck_shihan} for a fit) whereas for
the measurements in \cite{mantia, mantia14} larger values of $w_N$ are relevant,
depending on the scale probed.
\\

\begin{figure}
\includegraphics[width=2.7in]{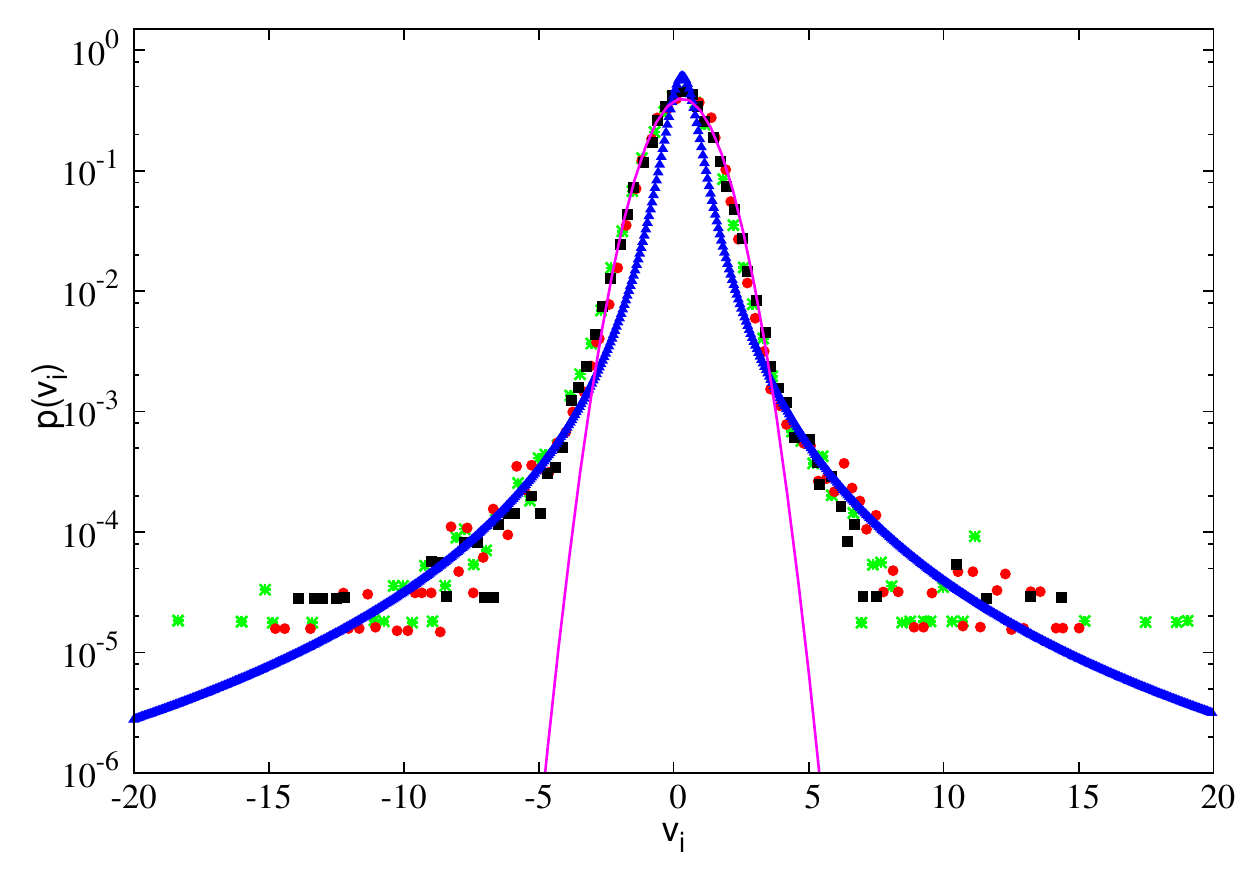}
\caption{Experimental data of La Mantia \textit{et al.} \cite{mantia} and a fit using eq.~(\ref{eq188}), (\ref{max}) and (\ref{gauss})
  with $\beta_0=4.5$ and  $c=0.3$ for the tail part (blue solid line), as well as a Gaussian with
  mean 0.3 and variance 1 for the central part.}
\label{vel}
\end{figure}

The PDFs of velocities of tracer particles measured in \cite{mantia14, mantia}
have power law tails but superimposed there is also a near-Gaussian distribution in the central part.
These measurements agree well with the prediction of our S/N/S model with strongly conditioned
$w_S|_v$ and $w_N|_v$, leading to formula (\ref{max}).
An excellent agreement is
obtained, as shown in Fig.\ref{vel}.
The critical velocity $v_c$ (in units of the standard deviation) is about $v_c \approx 4$,
whereas for the data of Paoletti et al. \cite{ref11} it is close to zero (see fit
in Fig.~1 in \cite{beck_shihan}),
meaning that in those measurements the particle is much more frequently in the S-phase.


\begin{figure}
\includegraphics[width=2.7in]{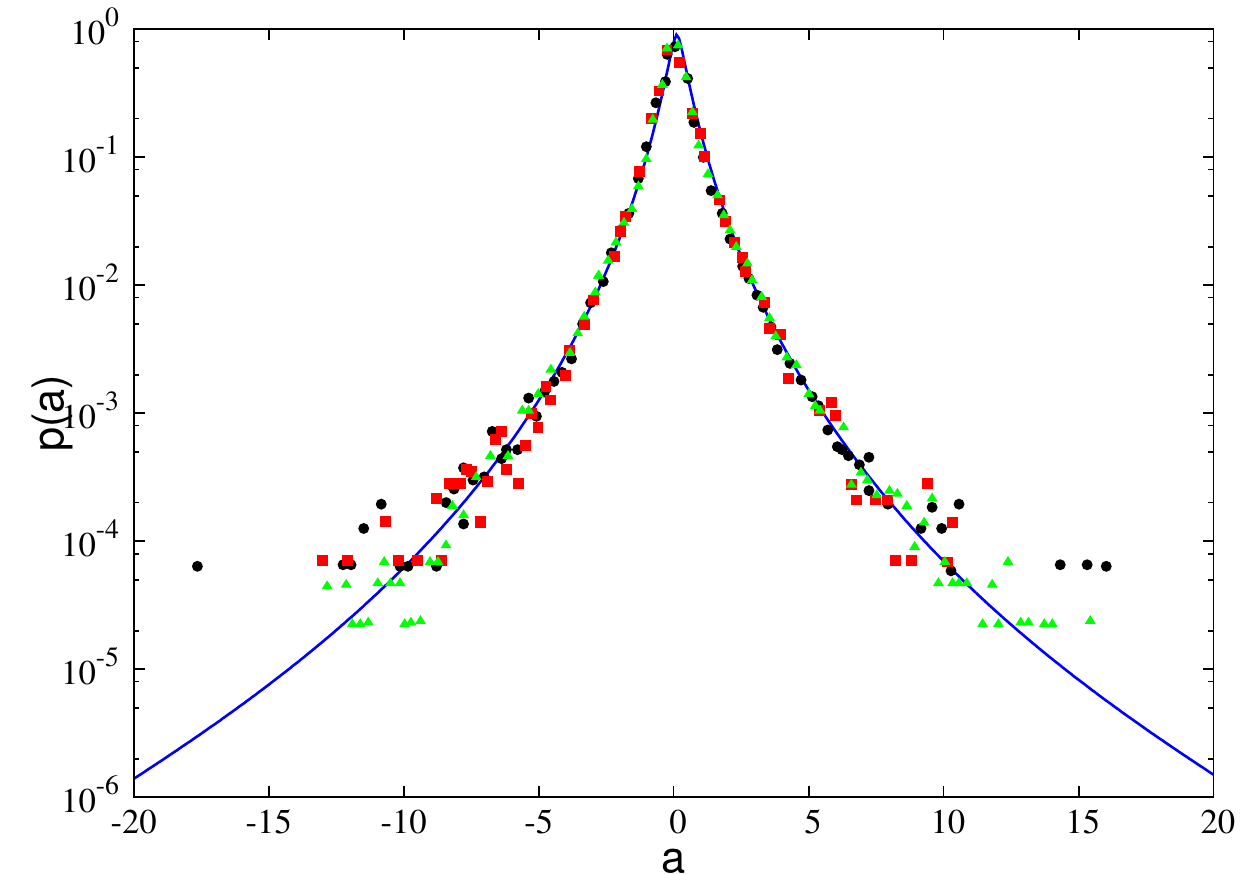}
\caption{ Experimental measurements of histograms of accelerations (in units
 of the standard deviation) obtained by La Mantia \textit{et al.}
 \cite{mantia}, and the Lognormal Superstatistics distribution given by
 Eq.(\ref{log}) with $s^2=2.2$, $b=e^{\frac{1}{2}s^2}$, $c=0.12$.}
\label{acc}
\end{figure}

Let us now move on from Lagrangian velocities
to Lagrangian accelerations. La Mantia \textit{et al.} \cite{mantia}
 also measured the probability density function of vertical particle accelerations
 in the quantum turbulent flow, see also \cite{mantia-new}
 for a recent update. They report that the tails of the density function become more pronounced
  as the temperature decreases and as the heat flux of the thermal counterflow increases.
  Again we expect an interplay between a different statistics for the normal phase N and
  the superfluid phase S of the embedded test particle. However,
  for the acceleration statistics the normal phase N will
  play a more important role than the phase S since the frictionless superfluid does not
  exhibit strong forces on the test particle.
  The cascade of energy dissipation is also similar in quantum turbulence as
  compared to classical turbulence \cite{salort}, so that the acceleration environment
  of a Lagrangian test particle is quite similar to the case of classical turbulence,
  at least on scales of the same order of magnitude as the average distance between
  quantized vortices.
  Hence, we basically expect a similar acceleration statistics
  in the quantum turbulent flow as for classical Lagrangian turbulence since quantum
  effects on the acceleration are small.
  For the acceleration statistics we may thus essentially consider superstatistical models than have been
  successfully applied before to reproduce acceleration statistics in classical turbulent flow,
  such as \cite{ref4}.
  These are based on lognormal superstatistics \cite{bcs}.
  The prediction of these types of models is that the PDF of acceleration components
  is given by
\begin{equation}
 p(a)=\frac{1}{2\pi s} \int_0^\infty \beta^{-1/2} exp{\left[\frac{-\left(log \frac{\beta}{b}\right)^2}{2 s^2}\right]} e^{-\frac{1}{2} \beta (a-c)^2} d\beta
\label{log}
\end{equation}
The above probability density function given by Eq.~(\ref{log}) is in excellent agreement with the experimental
measurements of La Mantia \emph{et al.} \cite{mantia}, as shown in Fig.~\ref{acc}.

For a more detailed understanding it is important to measure not only
histograms of velocities and accelerations, but also 2-point and higher correlation functions.
Depending on time scale, and the spatial scale probed by the test particles,
as well as the temperature of the quantum liquid,
the symbol sequences SNSSS... will have memory effects, as mentioned before.
Under simple model assumptions our approach
allows us to calculate the two-point correlation function analytically.
First, let us slightly generalize eq.~(\ref{eq1}) by introducing a memory
kernel. This allows for non-Markovian effects in the quantum liquid,
represented by memory effects in the symbol sequences.
The more general local dynamics including memory reads
\begin{equation}
 \dot {\boldsymbol{v}}=-\gamma\boldsymbol{v}
 +\alpha \int_{-\infty}^t dt'e^{-\eta(t-t')}\boldsymbol{v}
 +\omega \big[\boldsymbol e(t) \times \boldsymbol v\big] +\sigma (t)\boldsymbol{L}(t).
\label{ge}
\end{equation}
Taking $\eta \rightarrow \infty$  in Eq.~(\ref{ge}), one can obtain the previous local dynamics without
memory kernel. Equivalently, the  same dynamics can be obtained by putting $\alpha=0 $. The dynamics described
by Eq.~(\ref{ge}) without  rotational term has been studied by Van der Straeten \emph{et al} \cite{straeten-beck} and it was found
to be very useful to reproduce experimental data of turbulent Taylor-Couette flow.\\

For simplicity, let us first consider the direction of rotation to be represented by
 $ \boldsymbol e=(0,0,1)$. Then Eq.~(\ref{ge}) reduces to
\begin{equation}
\begin{split}
 \dot v_x&=-\gamma {v}_x-\omega {v}_y+\alpha \int_{-\infty}^t dt' e^{-\eta(t-t')} v_x(t')+ \sigma {L}_x(t)\\
\dot v_y&=-\gamma v_y+ \omega v_x+\alpha \int_{-\infty}^t dt' e^{-\eta(t-t')} v_y(t')+\sigma L_y(t)\\
\dot v_z&=-\gamma v_z+\alpha \int_{-\infty}^t dt' e^{-\eta(t-t')} v_z(t')+\sigma L_z(t)\\
\end{split}
\label{gec}
\end{equation}
Introducing a complex variable by defining $ u(t)=v_x(t)+iv_y(t)\;\mbox{and}\; g(t)=L_x(t)+iL_y(t)$,
the first two equations of Eq.~(\ref{gec}) can be written as
\begin{equation}
\begin{split}
 \frac{du}{dt}& =-(\gamma-i \omega)u+\alpha \int_{-\infty}^t dt' e^{-\eta(t-t')} u(t')+\sigma g(t)\\
\end{split}
\label{ged}
\end{equation}

To ease our calculations we introduce another variable by defining $\; Y=\alpha \int_{-\infty}^t dt' e^{-\eta(t-t')} u(t')$ and
then rewrite Eq.~(\ref{ged}) as a two-dimensional system of differential equations as follows
\begin{equation}
\begin{split}
 \dot u=&-(\gamma-i \omega)u+ Y +\sigma g(t)\\
\dot Y=&-\eta Y+\alpha  u(t)
\end{split}
\label{get}
\end{equation}
This form of the equations is helpful to perform numerical simulation as well as to calculate
the stationary distribution of the system.\\
In two-dimensional
classical turbulent point vortex dynamics correlation behaviour that is well fitted by
a sum of two exponentials has been observed (Fig.~4 in \cite{pasquero2001}).
 Correlation functions in three-dimensional quantum turbulence
 are constrained by the quantization condition and expected to be of
similar complexity as in the two-dimensional point vortex case.
One may thus conjecture that similar shapes of correlation functions are relevant. This we will derive now.

Using Fourier transformations in Eq.~(\ref{get}), we obtain
\begin{equation}
\begin{split}
&\langle u^{\ast}(t) u(0)\rangle=\\
& \oint  \frac{\sigma^2(\eta^2+ z^2)\;dz e^{iz t}}{\pi\left[(\gamma -i\omega-iz)(\eta-iz)-\alpha\right]\; \left[ (\gamma +i\omega+iz)(\eta+iz)-\alpha \right]}
\end{split}
\label{gei}
\end{equation}
Here, $u^*(t)$ is the complex conjugate of $u(t)$. As a special case for $\alpha=0$ and $\omega=0$ one can easily obtain the correlation
function $C_v(t)=\langle v_x(t+\tau)v_x(t)\rangle$ from Eq.~(\ref{gei}) as  $C_v(t)=\frac{\sigma^2}{2\gamma}  e^{-\gamma t}$.
 Similarly for $\alpha=0$  one obtains $C_v(t)=\frac{\sigma^2}{2\gamma}  e^{-\gamma t}\;\cos(\omega t)$. \\

For non zero $\alpha$ and $\omega$ a longer
calculation to evaluate the integral given by Eq.~(\ref{gei}) yields the following form of the correlation function:
\begin{equation}
\begin{split}
C_v(t)&=\frac{\sigma^2 e^{-Dt}}{2\sqrt{\mathcal{R}}QD}\Bigg[ \Pi \cos{(Ct-\theta)}+\Upsilon \sin{(Ct-\theta)}\Bigg]\;\;\;\;\;\;\;\;\;\;\;\;\;\;\;\;\\
&+\frac{\sigma^2 \;e^{-D't}}{2\sqrt{\mathcal{R}}QD'}\Bigg[ \Pi' \cos{(C't-\theta)}-\Upsilon' \sin{(C't-\theta)}\Bigg]\\
\end{split}
\label{crf2}
\end{equation}
Here the parameters are given as follows:
\begin{equation*}
 \begin{split}
\mathcal{R}&=\left[{\left[\omega^2-(\gamma-\eta)^2-4\alpha\right]^2}+4\omega^2(\gamma-\eta)^2\right]^{1/2}\\
\theta&=\frac{1}{2}\tan^{-1}\left[\frac{2\omega(\gamma-\eta)}{\omega^2-(\gamma-\eta)^2-4\alpha}\right]\\
C&=\frac{1}{2}\left[\omega+\sqrt{\mathcal{R}}\cos(\theta)\right] \\
D&=\frac{1}{2}\left[\gamma+\eta+\sqrt{\mathcal{R}}\sin(\theta)\right]\\
C'&=\frac{1}{2}\left[\omega-\sqrt{\mathcal{R}}\cos(\theta)\right] \\
D'&=\frac{1}{2}\left[\gamma+\eta-\sqrt{\mathcal{R}}\sin(\theta)\right]\\
Q&=(\gamma+\eta)^2+\mathcal{R} \cos^2(\theta)\\
\Phi&=\eta^2+C^2-D^2\\
\Phi'&=\eta^2+C'^2-D'^2\;\;\\
\Pi&=\Phi \sqrt{\mathcal{R}} \cos(\theta)+2CD(\gamma+\eta)\\
 \Pi'&=\Phi' \sqrt{\mathcal{R}} \cos(\theta)-2C'D'(\gamma+\eta)\\
\Upsilon&=\Phi(\gamma+\eta)-2CD\sqrt{\mathcal{R}}\cos(\theta)\\
 \Upsilon'&= \Phi'(\gamma+\eta)+2C'D'\sqrt{\mathcal{R}}\cos(\theta)
 \end{split}
\end{equation*}

\begin{figure}
\begin{center}
\begin{tabular}{cc}
\includegraphics[height=2.0in]{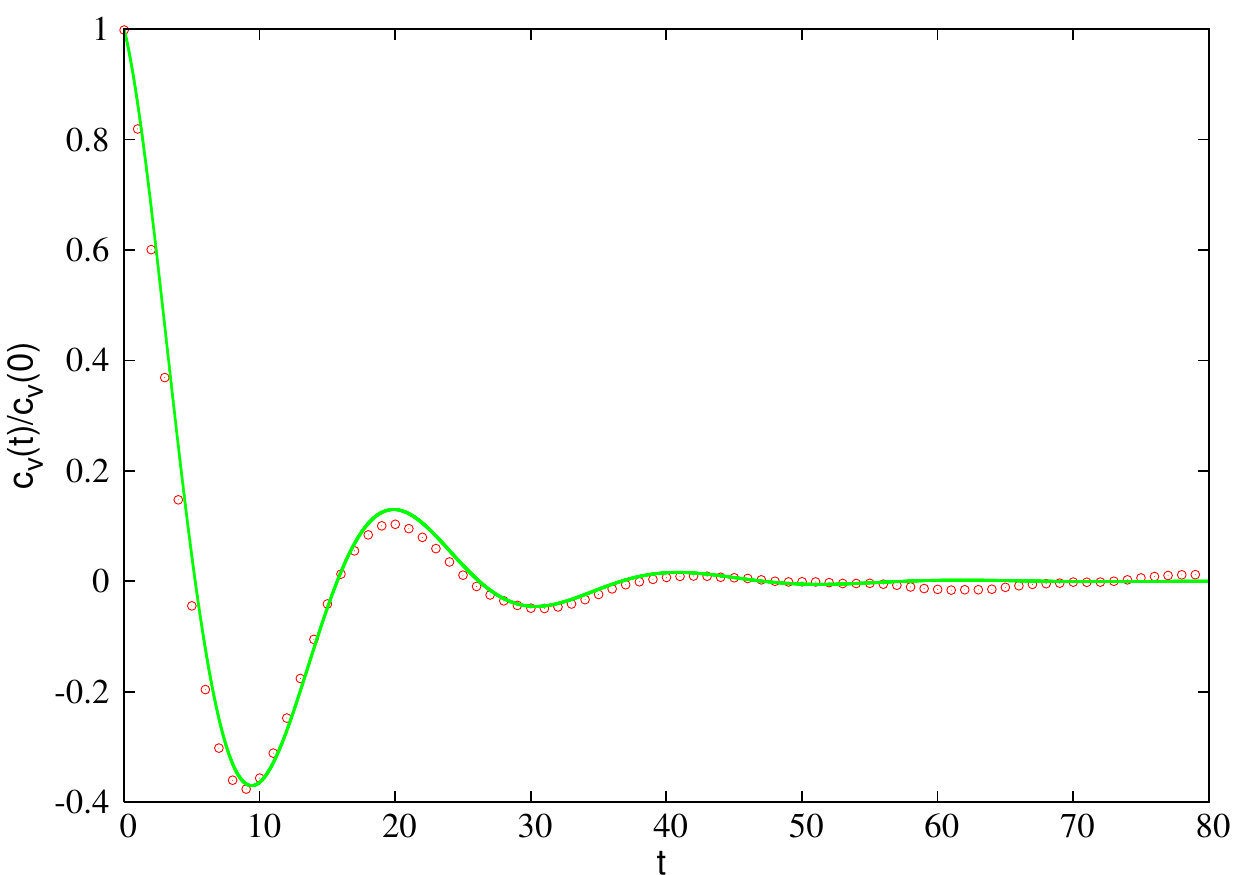}&
\end{tabular}
\end{center}
\caption{
Correlation function of a single velocity component as predicted by the generalized dynamics given
in Eq.~(\ref{crf2}) for $\gamma=0.1$, $\sigma=1.0$, $\omega=0.3$ and  $\alpha=0.0$ .
Red (data points) and green (solid lines) colours represent analytical and numerical results respectively.}
\label{alphazero}
\end{figure}

\begin{figure}
\begin{center}
\begin{tabular}{cc}
\includegraphics[height=2.0in]{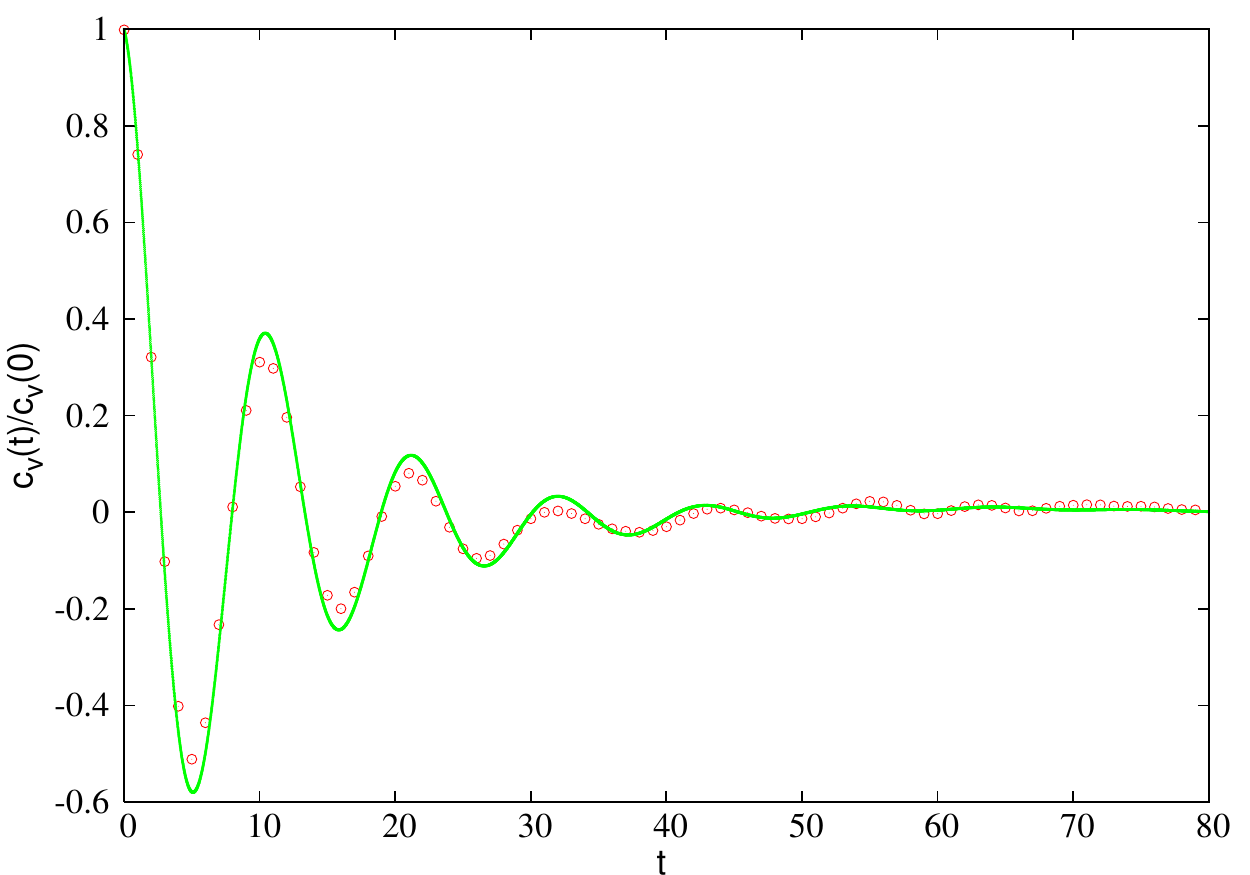}
\end{tabular}
\end{center}
\caption{Same as Fig.~3 but with memory effect, $\gamma=0.1$, $\sigma=1.0$, $\omega=0.5$ and $\alpha=-0.05$}
\label{alphaomega}
\end{figure}

The numerically simulated correlation function from the
Langevin dynamics for non-zero parameters
 $\omega$, $\gamma$, $\sigma$ and $\eta$ and the analytic result (\ref{crf2}) are plotted
in Figs.~\ref{alphazero} and \ref{alphaomega}.
As expected the results are in good agreement.  Similar shapes of autocorrelation functions of velocity
components are observed in atmospheric turbulence \cite{Reynolds2002,Reynolds2006,pasquero2001}.
Our results for the shape of correlation functions remain valid if $\sigma (t)$ fluctuates in a superstatistical way,
in this case $\sigma^2$ is simply replaced by its average $\langle \sigma^2 \rangle$. Similarly,
if different dynamical parameters are assumed in the S- and N-phases, one can average
the correlation function
over these.
It would be very interesting to check in future Lagrangian quantum turbulence measurements
whether shapes of correlation functions similar to those predicted in
Fig.~3 and 4 are observed.
\\ 

To summarize, in this paper we
have constructed a model for the dynamics of tracer particles in quantum turbulent flow, based
on a mixture of contributions from the superfluid (S) and the normal fluid (N). We discussed the
symbolic dynamics generated by the symbol sequences consisting of S and N.
We showed that the model generates velocity statistics of the tracer particle
that obeys a power law $p(v)\propto {v}^{-3}$ for large $v$,
together with a Gaussian core in the central region. The relative contribution of both densities
is determined by the statistics of the symbols S and N.
When the velocity is small
 the particle is more likely to be driven by the normal fluid.
 For very large velocities it is much more likely to be surrounded by superfluid and to
 be, in fact, close to a vortex core.
 Our analytic results are in very good agreement
with recent experimental data \cite{mantia14, mantia, ref11}.
We also showed that the acceleration statistics is well fitted by lognormal superstatistics,
 in a similar way as observed for classical Lagrangian turbulence \cite{ref4,bodenschatz}.\\
To better understand memory effects in the symbolic dynamics, we extended our
model by introducing a memory kernel so that the two-point correlation function of
the velocity component can capture more complicated dynamics.
We presented analytic results for the correlation function.
The general correlation function was shown to decay as a sum of two exponentials
with oscillatory behaviour.
 Future experimental measurements of correlation
functions of tracer particles could help to single out the optimum class of stochastic models,
and to better understand the properties of the symbolic dynamics
of quantum turbulent flow as a function of the external
control parameters.

\end{document}